\documentclass[prd,aps,superscriptaddress,showpacs,amsmath,
twocolumn,amssymb]{revtex4}
\usepackage{graphicx}

\newcommand{\errsp}{\,\,}

\newcommand{\tauint}{\tau_{\mathrm{int}}}

\newcommand{\xLS}{x_{\mathrm{LS}}}

\newcommand{\abovepoints}{(blue)\ }
\newcommand{\belowpoints}{(red)\ }

\begin{document}

\title{Discretisation effects in the topological susceptibility 
in lattice QCD}

\author{A. \surname{Hart}}
\affiliation{School of Physics, University of
Edinburgh, King's Buildings, Edinburgh EH9 3JZ, U.K.}

\collaboration{UKQCD and QCDSF Collaborations}
\noaffiliation

\begin{abstract}
  We study the topological susceptibility, $\chi$, in QCD with two
  quark flavours using lattice field configurations that have been
  produced with an $O(a)$--improved clover quark action. We find clear
  evidence for the expected suppression at small quark mass, $m_q$,
  and examine the variation of $\chi$ with this mass and the lattice
  spacing, $a$. A joint continuum and chiral extrapolation yields good
  agreement with theoretical expectations as $a,m_q \to 0$. A moderate
  increase in autocorrelation is observed on the more chiral
  ensembles, but within large statistical errors. Finite volume
  effects are negligible for Leutwyler--Smilga parameter $\xLS \gtrsim
  10$, and no evidence for a nearby phase transition is observed.
\end{abstract}

\preprint{Edinburgh 2003/23}

\pacs{11.15.Ha, 
      12.38.Gc  
     }

\maketitle

\section{Introduction}
\label{sec_introduction}

QCD is distinguished from the constituent quark models of hadronic
physics by the presence of light sea quarks in the vacuum and one of
the principal goals of lattice field theory is to provide an accurate
theoretical description of their effects.

The topological charge, $Q$, and its associated susceptibility,
$\chi$, are especially sensitive to the presence and properties of the
sea quarks. Chiral perturbation theory describes their variation with
the quark mass, $m_q$. By comparing lattice Monte Carlo measurements
with this expectation, we may understand how near we are to achieving
the above aim.

Initial work on unimproved sea quark formulations (Wilson
\cite{Bali:2001gk}
and staggered
\cite{Bitar:1991wr,Kuramashi:1992vq,Kuramashi:1993mv,Alles:1999kf,
Alles:2000cg,Hasenfratz:2001wd})
showed only weak evidence for the expected chiral ($m_q \to 0$)
suppression of $\chi$ relative to the quenched ($m_q \to \infty$)
theory at an equivalent lattice spacing, $a$ (see
\cite{Hart:2000wr,Durr:2001ty,Hart:2001pj}
for reviews). A clearer chiral reduction was seen for improved fermion
formulations, probably as a result of better chiral symmetry at finite
lattice spacing
\cite{Hasenfratz:2001wd}.
${\cal O}(a)$--improved Sheikholeslami--Wohlert (SW, or ``clover'')
Wilson fermions showed a strong sea quark effect in a study at
quasi--fixed lattice spacing, $a \simeq 0.1~\mathrm{fm}$
\cite{Hart:1999hy,Hart:2000wr,Hart:2000gh,Allton:2001sk,Hart:2001pj,
Hart:2001tv};
mean field improved SW quarks also showed an effect
\cite{AliKhan:2001ym}.
Improved staggered fermion studies have also shown clear suppression
of $\chi$ over a range of lattice spacings $0.09 \lesssim a \lesssim
0.12~\mathrm{fm}$
\cite{Hasenfratz:2001wd,Bernard:2002sa,Bernard:2003gq}.
Studies using lattice sea quark formulations with chiral properties
more comparable to those of the continuum have so far been limited to
two dimensional models
\cite{Durr:2003xs}.

In the light of the improved staggered results, it is interesting to
examine the effects of finite lattice spacing on the topological
susceptibility from ${\cal O}(a)$--improved SW quarks. This is useful
not only from the theoretical viewpoint, but also when comparing the
relative costs of producing independent field configurations in Monte
Carlo simulations (the topological modes are expected to be amongst
the slowest to decorrelate).

In this paper we extend the analysis of two flavour ${\cal
  O}(a)$--improved SW results at $a \simeq 0.1~\mathrm{fm}$ (included
in this work with due acknowledgment) to cover a comparable range of
lattice spacings to the improved staggered fermion study, using
lattice QCD ensembles produced by the UKQCD and QCDSF collaborations.
Moving to small quark masses (the lightest yet produced for
Wilson--like fermions), we find strong evidence for the chiral
suppression of $\chi$. Finite volume effects are shown to be under
control, and no strong autocorrelation is seen for the topological
charge. We perform a joint continuum chiral and continuum
extrapolation, and find good agreement with theoretical expectations.

The structure of this paper is as follows: in
Section~\ref{sec_top_susc} we review the theoretical expectations for
the chiral variation of the topological susceptibility in the
continuum and on the lattice. We describe our measurements in
Section~\ref{sec_meas}, and discuss their variation with $m_q$, $a$
and the lattice volume. We conclude with a discussion in
Section~\ref{sec_summary}.

\section{The topological susceptibility}
\label{sec_top_susc}

Four--dimensional, Euclidean gauge fields can be classified by an
integer--valued topological charge,
\begin{equation}
Q = \int d^4x \frac{1}{2} \varepsilon_{\mu \nu \sigma \tau} 
F_{\mu \nu}(x) F_{\sigma \tau}(x)
\in 
\mathbb{Z} \; .
\end{equation}
In the absence of a $\theta$--term, $\left \langle Q \right \rangle =
0$ and the topological susceptibility, $\chi = \left\langle Q^2
\right\rangle / V$, has a finite, non--zero limit as the volume $V \to
\infty$.

\begin{table*}[tb]

\caption{\label{tab_ensembles}The ensembles including the volume
$L^3T$ in lattice units, the number of configurations analysed, their
separation in HMC trajectories, physical parameters and various
measures of the physical volume.}
\begin{ruledtabular}
\begin{tabular}{lllllllrrr}
\multicolumn{1}{c}{label} &
\multicolumn{1}{c}{$\beta$} &
\multicolumn{1}{c}{$\kappa_s $} &  
\multicolumn{1}{c}{$L^3 T$} &
\multicolumn{1}{c}{$N_{\text{conf}}$, sepn} &
\multicolumn{1}{c}{$r_0$} & 
\multicolumn{1}{c}{$(r_0 m_\pi)^2$} & 
\multicolumn{1}{c}{$L/r_0$} & 
\multicolumn{1}{c}{$Lm_\pi$} &
\multicolumn{1}{c}{$\xLS$} \\
\hline
$q_1$ &  5.20  &  0.13420  &  $16^3 32$  &  514, 10 & 
        4.077 \errsp(31) & 5.67 \errsp(7) & 3.9 &  9.3 & 150 \\
$u_2$ &  5.20  &  0.13500  &  $16^3 32$  &  789, 10 & 
        4.754 \errsp(40) & 3.87 \errsp(5) & 3.4 &   6.6 &  55 \\
$u_3$ &  5.20  &  0.13550  &  $16^3 32$  &  830, 10 & 
        5.041 \errsp(40) & 2.15 \errsp(3) & 3.2 &   4.6 &  24 \\
$u_4$ &  5.20  &  0.13565  &  $16^3 32$  &  280, 10 & 
        5.246 \errsp(51) & 1.67 \errsp(4) & 3.0 &   3.9 &  16 \\
$u_5$ &  5.20  &  0.13580  &  $16^3 32$  &  279, 10 & 
        5.320 \errsp(50) & 1.22 \errsp(4) & 3.0 &   3.3  & 11 \\
\hline
$q_6$ &  5.25  &  0.13460  &  $16^3 32$  &  194, 10 & 
        4.737 \errsp(21) & 5.44 \errsp(4) & 3.4 &   7.9 &  79 \\
$u_7$ &  5.25  &  0.13520  &  $16^3 32$  &  822, 10 & 
        5.138 \errsp(45) & 3.90 \errsp(5) & 3.1 &   6.1 &  41 \\
$q_8$ &  5.25  &  0.13575  &  $24^3 48$  &   91, 10 & 
        5.430 \errsp(60) & 1.99 \errsp(4) & 4.4 &   6.2 &  85 \\
\hline
$u_9$ &  5.26  &  0.13450  &  $16^3 32$  &  415, 10 & 
        4.708 \errsp(52) & 5.73 \errsp(9) & 3.4 &  8.1 & 85 \\
\hline
$u_{10}$ &  5.29  &  0.13400  &  $16^3 32$  &  397, 10 & 
        4.813 \errsp(45) & 7.70 \errsp(11) & 3.3 &   9.2 & 105 \\
$q_{11}$ &  5.29  &  0.13500  &  $16^3 32$  &  587, 5 & 
        5.227 \errsp(37) & 4.84 \errsp(5) & 3.1 &   6.7 &  47 \\
$q_{12}$ &  5.29  &  0.13550  &  $12^3 32$  &  834, 5 & 
        5.756 \errsp(33) & 3.53 \errsp(3) & 2.1 &   3.9 &  10 \\
$q_{13}$ &  5.29  &  0.13550  &  $16^3 32$  &  911, 5 & 
        5.560 \errsp(30) & 3.30 \errsp(3) & 2.9 &   5.2 &  25 \\
$q_{14}$ &  5.29  &  0.13550  &  $24^3 48$  & 405, 5 & 
        5.566 \errsp(20) & 3.30 \errsp(2) & 4.3 & 7.8 & 127 \\
\hline
$q_{15}$ &  5.40  &  0.13500  &  $24^3 48$  & 653, 2 & 
        6.088 \errsp(32) & 6.03 \errsp(5) & 3.9 &   9.7 & 162 
\end{tabular}
\end{ruledtabular}
\end{table*}

As the mass of the sea quarks is reduced the topological
susceptibility is suppressed, the leading order behaviour being
\cite{Crewther:1977ce,DiVecchia:1980ve,Leutwyler:1992yt}
\begin{equation}
\chi(m_\pi^2) = \frac{(f_\pi m_\pi)^2}{4}
\label{eqn_chi_chi}
\end{equation}
for two degenerate flavours, with corrections at ${\cal O}(m_\pi^4)$.
This relation should hold when $V$ is large enough that chiral
symmetry is not restored, i.e.
\begin{equation}
\xLS \equiv m_q \Sigma V \simeq (f_\pi m_\pi)^2 V \gg 1 \; ,
\label{eqn_xLS}
\end{equation}
where $\Sigma$ is the chiral condensate
\cite{Leutwyler:1992yt}
\footnote{ By the ``pion'' we mean the lightest pseudoscalar meson
  with valence quarks of the same mass as the two degenerate sea
  quarks flavours The PCAC relation predicts $m_q \propto m_\pi^2$ and
  we shall use the latter as a measure of the quark mass.}.

As $m_q \to \infty$, $\chi$ approaches the (constant) quenched value,
$\chi^{\mathrm{(qu)}}$, from below. Higher order corrections to
Eqn.~(\ref{eqn_chi_chi}) thus introduce a negative curvature at some
intermediate quark mass. Various interpolating ans\"{a}tze between the
chiral and quenched regimes have been proposed
\cite{Leutwyler:1992yt,Hart:2000wr,Durr:2001ty}.
In this paper we shall focus on one particular form, the large-$N_c$
form
\cite{Leutwyler:1992yt}, 
as motivated for SU(3) by D\"{u}rr
\cite{Durr:2001ty}:
\begin{equation}
r_0^4 \chi = \frac{c_3 c_0 (r_0 m_\pi)^2}{c_3 + c_0(r_0 m_\pi)^2} 
~~ \mathrm{with} ~
\begin{array}[t]{l}
c_0 = (r_0 f_\pi)^2/4 \; ,\\
c_3 = r_0^4 \chi^{\mathrm{(qu)}} \; .
\end{array}
\label{eqn_nlge}
\end{equation}
Assuming the physical values of $r_0 = 0.5~\mathrm{fm}$
\cite{Sommer:1994ce}
and $f_\pi = 93~\mathrm{MeV}$, then $c_0 = 0.0139$. This formula has
been seen to work well in QCD
\cite{Hart:2000wr,Durr:2001ty,Hart:2001pj}
at fixed lattice spacing.

\subsection{Lattice QCD}

Eqn.~(\ref{eqn_chi_chi}) also holds for lattice QCD at finite lattice
spacing for fermions with sufficiently good chiral properties
\cite{Chandrasekharan:1998wg,Giusti:2001xh}.
In general, however, the lattice topological susceptibility is related
to the continuum by additive and multiplicative renormalisation
factors, $Z$ and $M$:
\begin{equation}
r_0^4 \chi^{\mathrm{(lat)}} = Z^2 r_0^4 \chi^{\mathrm{(cont)}} + M
\; .
\end{equation}
These competing factors are in general functions of the lattice
spacing and the quark mass (partly implicitly through the coupling,
$g^2$). The upshot of this is that discretisation effects act to
reduce $\chi$ below its continuum limit at large $m_q$, but to
increase it in the chiral limit. A recent discussion can be found in
\cite{Bernard:2003gq}.
Formally, before examining the sea quark mass dependence, we should
first perform a continuum extrapolation of $\chi$ at fixed $m_q$ to
remove the effect of these renormalisation factors. This is, however,
not possible with the ensembles we have available (especially if we do
not admit interpolation in $m_q$ of measurements at finite $a$). To
attempt a full analysis of our data we must assume that the continuum
and chiral limits commute sufficiently well that we can either perform
the $m_q \to 0$ limit first at fixed $a$, or that we can carry out a
joint chiral and continuum extrapolation using a single formula.

The first approach was used in 
\cite{Hart:2000wr,Hart:2001pj},
fitting Eqns.~(\ref{eqn_chi_chi}) and~(\ref{eqn_nlge}) to data at $a
\simeq 0.1~\mathrm{fm}$. Constraints of simulation mean that this
study cannot contribute many ensembles at comparably light sea quark
mass but different~$a$. We cannot, therefore, extend the above method
by carrying out chiral fits at other lattice spacings so we can look
at discretisation effects in the fit parameters. Instead we follow the
second approach. Allowing $c_{0,3}$ to have a leading order lattice
spacing correction (quadratic for the lattice action considered here)
\begin{equation}
c_0(r_0) = c_{00} + \frac{c_{01}}{r_0^2} \; ,
~~~~
c_3(r_0) = c_{30} + \frac{c_{31}}{r_0^2} \; ,
\end{equation}
in Eqn.~(\ref{eqn_nlge}), we obtain:
\begin{widetext}
\begin{equation}
r_0^4 \chi =
\frac{c_{30} c_{00} (r_0 m_\pi)^2}{c_{30} + c_{00}(r_0 m_\pi)^2} + 
\frac{1}{r_0^2} \left(
k_1 + 
\frac{(r_0 m_\pi)^2}{(c_{30} + c_{00}(r_0 m_\pi)^2)^2}
\left(c_{01}c_{30}^2 + c_{31}c_{00}^2 (r_0 m_\pi)^2 \right)
\right) \; .
\label{eqn_nlge_a}
\end{equation}
\end{widetext}
We have added a general constant, $k_1 \ge 0$, to allow $\chi$ to have
a non-zero chiral limit at finite lattice spacing, as might occur in
the absence of exact zero modes in the lattice Dirac operator
\cite{Hart:2000wr}.
A check of our results is consistency of the fitted parameters with
the results for $r_0 \simeq 5$ from
\cite{Hart:2000wr,Hart:2001pj}.
\section{Measurements}
\label{sec_meas}

We have calculated the topological charge and susceptibility using
ensembles generated by the UKQCD
\cite{Allton:2001sk,Irving:2002fx,Ukqcd:2003}
and QCDSF 
\cite{Booth:2001qp,Khan:2003cu}
collaborations. Details of these ensembles are given in
Table~\ref{tab_ensembles} (taken in the main from
\cite{Khan:2003cu}).
The lattice spacings lie in the range $0.08 \lesssim a
\lesssim 0.12~\mathrm{fm}$. The SU(3) gauge fields are governed by the
Wilson plaquette action, with two flavours of SW fermions. The
improvement parameter, $c_{\text{sw}}$, has been chosen so that the
leading order discretisation errors vary quadratically with the
lattice spacing
\cite{Jansen:1998mx}.
The exact Hybrid Monte Carlo (HMC) simulation algorithm is used, which
avoids finite step-size errors.

\begin{figure}[b]

\includegraphics[width=3in,clip]{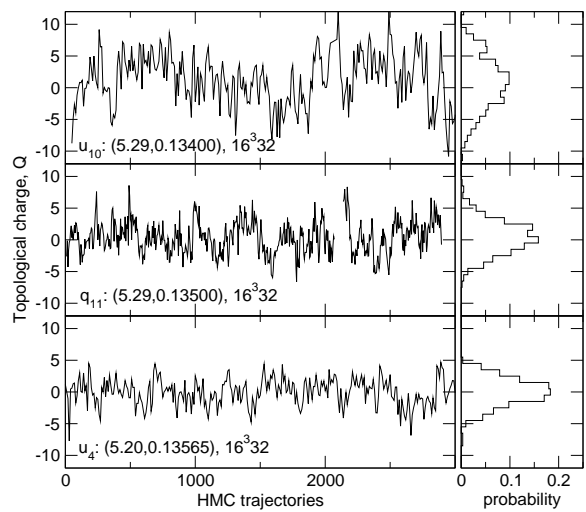}
\caption{\label{fig_thist_chiral}Partial time histories for the
topological charge for ensembles $u_{10,4}$ and $q_{10}$, with
normalised histograms of the topological charge.}
\end{figure}

We measure $Q$ using the method of
\cite{Hart:2000wr}:
ten cooling sweeps are applied using the Wilson gauge action. The
cooling action has relatively little effect on the topological
susceptibility
\cite{Bernard:2003gq}.
Ten cools strikes a good balance between adequate suppression of these
ultraviolet dislocations and excessive destruction of the long range
topological structure
\cite{Hart:2000wr,Bernard:2003gq}.
The choice of operator does not matter in the continuum limit, at
least for the quenched theory
\cite{Teper:1999wp},
and a reflection--symmetrised ``twisted plaquette'' lattice
topological charge operator is used. In Fig.~\ref{fig_thist_chiral} we
show the variation of the topological charge as a function of Monte
Carlo simulation time for three ensembles over a range of sea quark
mass. The histograms of $Q$ (using unit width bins, although in the
rest of the analysis the topological charge is not rounded to the
nearest integer value) show good agreement with the expected Gaussian
form. Although $\langle Q \rangle = 0$ within statistical errors, we
opt (for consistency with earlier studies) to subtract terms in
$\langle Q \rangle$ from $\chi$. Both this and the choice of whether
to round $Q$ to integer values affect $\chi$ by much less than one
standard deviation.

This algorithm for measuring $\chi$ has already been studied in the
quenched theory (the Wilson gauge action), and continuum extrapolation
yields $c_{30} = 0.065~(3)$ and $c_{31} = -0.28~(4)$
\cite{Hart:2001pj}.
\begin{table}[b]

\caption{\label{tab_results}The integrated autocorrelation estimates
for the topological charge, and topological susceptibility.}
\begin{ruledtabular}
\begin{tabular}{lll}
\multicolumn{1}{c}{label} &
\multicolumn{1}{c}{$\tauint$} & 
\multicolumn{1}{c}{$r_0^4 \chi$} \\
\hline
$q_1$ &  11 \errsp(2) & 0.0494 \errsp(32) \\
$u_2$ &  16 \errsp(3) & 0.0445 \errsp(30) \\
$u_3$ &  30 \errsp(13) & 0.0310 \errsp(38) \\
$u_4$ &  26 \errsp(8) & 0.0248 \errsp(28) \\
$u_5$ &  29 \errsp(15) & 0.0230 \errsp(25) \\
\hline
$q_6$ &  12 \errsp(1) & 0.0451 \errsp(32) \\
$u_7$ &  19 \errsp(4) & 0.0392 \errsp(28) \\
$q_8$ &  92 \errsp(16) & 0.0302 \errsp(68) \\
\hline
$u_9$ &  14 \errsp(4) & 0.0433 \errsp(36) \\
\hline
$u_{10}$ &  125 \errsp(40) & 0.0683 \errsp(68) \\
$q_{11}$ &  20 \errsp(6)  & 0.0369 \errsp(42) \\
$q_{12}$ &  78 \errsp(35) & 0.0306 \errsp(52) \\
$q_{13}$ &  22 \errsp(8) & 0.0276 \errsp(33) \\
$q_{14}$ &  25 \errsp(15) & 0.0356 \errsp(57) \\
\hline
$q_{15}$ &  150 \errsp(22) & 0.0325 \errsp(87)
\end{tabular}
\end{ruledtabular}
\end{table}

\begin{figure}[b]

\includegraphics[width=2.5in,clip]{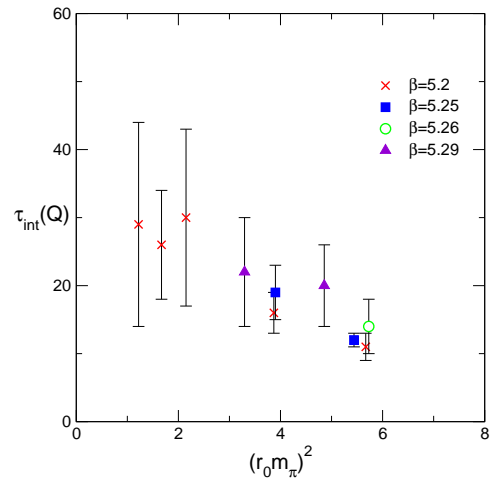}
\caption{\label{fig_tau_int}Integrated autocorrelation estimates
for topological charge on the $16^3 32$ lattice volumes,
labeled by gauge coupling, $\beta$.}
\end{figure}

The decorrelation of $Q$ is good (see Fig.~\ref{fig_thist_chiral}, and 
\cite{Hart:2000wr,Hart:2001pj,Ukqcd:2003}
for details of some other ensembles). More quantitatively, we measure
the integrated autocorrelation, $\tauint(Q)$, using an implementation
of the sliding window method
\cite{Madras:1988ei},
with results shown in Table~\ref{tab_results} (scaled to units of HMC
trajectories)
\footnote{Autocorrelations are notoriously hard to estimate in
realistic ensembles, and for consistency of analysis here there are
slight differences from earlier studies of the intermediate quark mass
ensembles
\cite{Hart:2000wr,Hart:2001pj}.
With this in mind, the results presented here have been compared to
those from a number of other algorithms for determining $\tauint$, and
show a good degree of robustness.}.
With the exceptions of ensembles $q_{8,12,15}$ and $u_{10}$, there is
weak evidence for a slow increase in $\tauint$ from ${\cal O}(10)$~HMC
trajectories for the least chiral ensembles to ${\cal
  O}(30)$~trajectories for the most chiral. We plot these data for the
$16^3 32$ ensembles in Fig.~\ref{fig_tau_int} (omitting the least
chiral point). The trend is no stronger for any particular gauge
coupling, $\beta$, but the statistical errors and available range of
$(r_0 m_\pi)^2$ make definitive statements difficult. Of the outliers,
we note that ensembles $q_{8,15}$ have only a limited number of
trajectories and, as we discuss later, $q_{12}$ is the smallest
physical volume studied.

\begin{figure}[b]

\includegraphics[width=2.5in,clip]{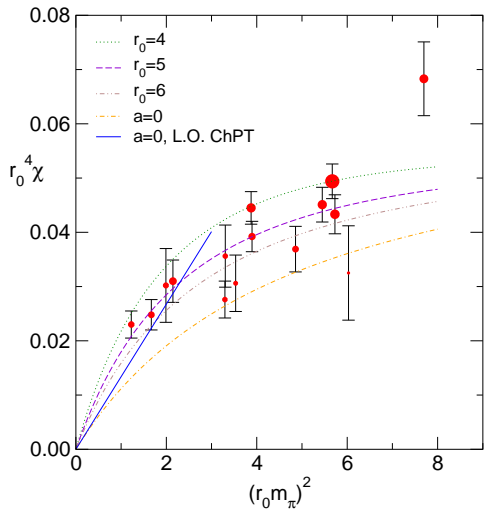}
\caption{\label{fig_r04_chi} The variation of the topological
  susceptibility. To guide the eye, the point size is proportional to
  $r_0^{-2}$.  The single fitted curve Eqn.~(\ref{eqn_nlge_a}) is
  shown for the continuum limit and lattice spacings representative of
  the plotted data. The expectation from continuum chiral perturbation
  theory at leading order is also shown.}
\end{figure}

Barring these possible exceptions, we have confidence that the
topological susceptibility may be estimated free from autocorrelation
effects and the statistical errors quoted are from a jack--knife
analysis using 10 bins. No increase in the statistical error estimates
for $\chi$ was found with larger bins, as expected when comparing
$\tauint$ to the bin size. Results for the topological susceptibility
are given in Table~\ref{tab_results} and plotted against quark mass in
Fig.~\ref{fig_r04_chi}.

\begin{table}

\caption[]{\label{tab_fits} Fits to
Eqn.~(\ref{eqn_nlge_a}). Numbers in italic face were fixed in the fit. Fit 2 was carried out with two different sets of initial values, as described in the text.}
\begin{center}
\begin{ruledtabular}
\begin{tabular}{llll}
parameter & Fit 1 & \multicolumn{2}{c}{Fit 2} \\
\hline
$c_{00}$ & 0.0135  (26) & 0.0036  (30) & 0.0283 (36) \\
$c_{01}$ & 0.2595  (229) & 0.1074  (250) & 1.453 (522) \\
$c_{30}$ & \textit{0.065} & \textit{0.065} & \textit{0.065} \\
$c_{31}$ & \textit{$-$0.28} & \textit{$-$0.28} & \textit{$-$0.28} \\
$k_1$ & \textit{0} & 0.0149  (51) & $-$0.028 (9) \\
$\chi^2/\mathrm{d.o.f.}$ & 1.043 & 0.980 & 0.721
\end{tabular}
\end{ruledtabular}
\end{center}

\end{table}
\subsection{Discretisation effects}

There is a significant suppression of the topological susceptibility
in the chiral limit. In Fig.~\ref{fig_r04_chi} we show the leading
order chiral expectation of Eqn.~(\ref{eqn_chi_chi}), using the
physical value for $f_\pi$.  There is approximate agreement for $(r_0
m_\pi)^2 \lesssim 3$, although we note that discretisation effects
tend to increase $\chi$ and exaggerate the range of agreement. We do
not see clear evidence for $\chi$ deviating from a rapid decrease with
$m_q$ at finite lattice spacing as in
\cite{Bernard:2003gq},
although critical slowing prevents exploration of sea quarks
comparably light with those simulated in that study.

To quantify the effects of lattice artifacts, it is useful to consider
$\chi$ over the full range of sea quark masses. Assuming that
Eqn.~(\ref{eqn_nlge_a}) holds, we can use it to perform a chiral
extrapolation of $\chi$ whilst allowing for (leading order)
discretisation effects. We exclude ensemble $u_{10}$, whose
statistical error is probably unreliable, and ensemble $q_{12}$ which
has a small volume (see on). Ensembles $q_{8,15}$ also have large
autocorrelation, and we remark that excluding them from the fitted
data had no statistically significant effect. We fit only the
dynamical results, with the quenched parameters fixed. In principle we
could avoid this and fit the quenched data as well, assigning all such
points an arbitrarily large value of $(r_0 m_\pi)^2$. Such fits do not
converge well, however. This was probably as a result of the data
being arranged as two distinct clusters in $(r_0 m_\pi)^2$.

\begin{figure}[b]
\includegraphics[width=3.3in,clip]{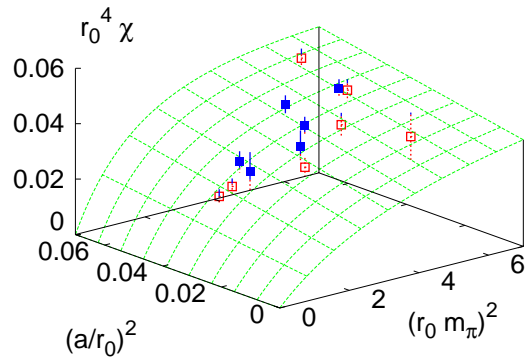}
\caption[]{\label{fig_3d} Combined chiral and continuum extrapolation
  of the topological susceptibility. Solid \abovepoints points and
  error bars denote data points lying above the fitted surface. Open
  \belowpoints points and dashed error bars denote data points lying
  below.}
\end{figure}

We show the fits to Eqn.~(\ref{eqn_nlge_a}) in Table~\ref{tab_fits},
with the $k_1 = 0$ results (Fit~1) plotted in Fig.~\ref{fig_r04_chi}
for a representative sample of lattice spacings, and in
Fig.~\ref{fig_3d}.

It is interesting to note that even for the heaviest sea quarks
simulated, the discretisation effects act to increase $\chi$. To see a
crossover to the quenched behaviour (where a decrease is seen), the
fits suggest we need $(r_0 m_\pi)^2 \gtrsim 25$. It is thus clear that
for all the ensembles the vacuum has been qualitatively altered from
that of the quenched theory by the presence of the sea quarks. As
expected, topological degrees of freedom appear to be more sensitive
to such changes than most hadronic observables.

For $r_0 = 5.0 \pm 0.2$ (the range studied in 
\cite{Hart:2000wr,Hart:2001pj}),
these fits suggest $c_0 = 0.024~(3)(1)(1)$, where the errors arise
from those on $c_{00}$, $c_{01}$ and the above range of $r_0$
respectively. This figure agrees very closely with $c_0 = 0.22~(7)$
obtained in the previous work for Eqn.~(\ref{eqn_nlge}), and gives
confidence in the consistency of our fits.

In addition to the errors arising from the statistical variation in
the data, there is also a systematic error arising from the
uncertainty in the quenched behaviour. We attempt to account for this
by repeating the fits with $c_{30}$ and $c_{31}$ varying by one
standard deviation in each direction: $c_{00}$ varies by $\pm 0.003$
under this, and it is clear that this is a sub--leading effect that
can be ignored.  

The fitted $c_{00} = 0.0135 ~ (26)$ corresponds to
$f_\pi = 91.6~(8.8)~\mathrm{MeV}$. This agrees surprisingly well both
with the experimental (and 2+1--flavour) QCD measurement
($93~\mathrm{MeV}$) and measurements by other methods on $N_f=2$ SW
actions ($\simeq 97.0~(0.3)~\mathrm{MeV}$ in these conventions,
rescaled from
\cite{Aoki:2002uc}).
The leading order lattice spacing correction to this increases
$f_\pi$, as expected. The size of this correction is consistent with
those obtained for other methods (see Fig.~35 of
\cite{Aoki:2002uc},
for instance, although the different gauge actions and operators will,
of course, lead to different lattice artifacts).

It is useful here to consider whether there is any evidence for $\chi$
having a non--zero chiral limit at finite $a$. In Fit~2 of
Table~\ref{tab_fits} we show the result lifting the restriction $k_1 =
0$. The fits are extremely unstable; the first column is the result of
starting with all free parameters being zero, whilst the second takes
the results of Fit~1 as the initial values. This instability probably
arises from the most chiral data points being clustered around $r_0 =
5$. We conclude that whilst we cannot rule out a non--zero chiral
limit, the data appears to be described more consistently when such a
term is not present. For the remainder of the paper we therefore
concentrate on the results from Fit~1.

\subsection{Finite volume effects}

In 
\cite{Bernard:2003gq}
the lattice volumes were sufficiently large that the topological
charge on subvolumes (of size $L^4$) varied independently; indeed a
reduction in statistical error on $\chi$ was achieved by exploiting
this self--averaging. We did not find the same in the this study, and
this prompts us to consider more carefully whether the topological
susceptibility exhibits finite volume contamination.

In general we expect finite volume effects to lead to a reduction in
$\chi$: a small lattice excludes large instantons. In the quenched
theory this restriction is negligible once $L/r_0 \gtrsim 2.5$
\cite{Lucini:2001ej}.
With dynamical sea quarks present, small volumes can, in addition, see
a restoration of chiral symmetry with the leading order chiral
variation now being
$
\chi \propto (m_q)^{N_f} \propto (m_\pi)^{2N_f}.
$
Avoiding this requires $\xLS \gg 1$, although the precise limit is not
known. What is also not clear is how the behaviour interpolates
between this and Eqn.~(\ref{eqn_xLS}) as we vary $\xLS$; we do not
address this in this study.

Finally, we need the lattice to be large in units of the pion
correlation length, $L m_\pi \gg 1$. Spectroscopy suggests $L m_\pi
\gtrsim 5.7$ is necessary to render the low-lying hadron states free of
finite volume effects
\cite{Allton:1998gi}.
This limit is, however, associated more with pion exchange than
phase structure and is probably unduly stringent for this study.

In Table~\ref{tab_ensembles} we show all three measures of FVEs. All
the above limits seem well satisfied and we do not expect, or see, any
sign of chiral symmetry restoration or significant finite volume
effects. Certainly we do not see the smaller volumes lying
consistently below the fitted curve.

The ensembles $q_{12,13,14}$ vary only in volume, and no strong
suppression of $\chi$ is seen as $V$ is reduced. Nonetheless, we
choose to exclude $q_{12}$, the smallest lattice, from the scaling
analysis as the volume is particularly small.

We can attempt to quantify the effects of finite volume by repeating
the fit with $k_1 = 0$, progressively excluding the ensembles with the
smallest $\xLS$ ($u_5$, $u_4$...) until the fits become unreliable.
The fit parameters remained consistent within statistical errors. We
conclude that finite volume effects are not significant at this level
of statistical accuracy if $\xLS \gtrsim 10$.

\section{Summary}
\label{sec_summary}

We have carried out a systematic study of the topological
susceptibility, $\chi$, in lattice QCD with two light quark flavours
over a range of lattice spacings and sea quark masses using ensembles
generated by the UKQCD and QCDSF collaborations. We find compelling
evidence for suppression of $\chi$ with decreasing quark mass. Finite
volume contamination of the results was not seen, and $\xLS \gtrsim
10$ is necessary to avoid such effects on the topological degrees
of freedom.

We observe a slight (and, in truth, statistically insignificant)
increase in the integrated autocorrelation of the topological charge,
$Q$, as the sea quark mass is reduced, although this effect is not
marked despite a critical slowing of the HMC algorithm
\cite{Irving:2002fx,Ukqcd:2003}. 
The autocorrelation times are, in general, small enough not to bias
the jack--knife estimates of the statistical errors on our data.

There have been suggestions that at $\beta \simeq 5.2$ the $N_f=2$
${\cal O}(a)$--improved SW action exhibits evidence for a nearby phase
transition
\cite{Hart:2001fp,Hart:2001tv,Sommer:2003ne}.
This transition affects most strongly observables with the quantum
numbers of the vacuum, and we do not expect (or find) a strong effect
on the topological susceptibility. It may, however, induce additional
mass dependence in $r_0$
\cite{Sommer:2003ne},
although we see no signal in $r_0^4 \chi$ (whose fourth power would
presumably magnify any such effect, distorting the data away from
Eqn.~(\ref{eqn_nlge_a})). It may also lead to large autocorrelation
\cite{Sommer:2003ne}.
Whilst a large integrated autocorrelation is seen for one ensemble at
large sea quark mass, this is at $\beta = 5.29$ which does not
immediately fit the hypothesis of
\cite{Sommer:2003ne}.
No unexpected trends were seen for $\beta=5.2$.

The chiral variation of the data was compared to a theoretically
motivated ansatz for the behaviour of $\chi$ over the full range of
sea quark masses. This was extended to allow for variation of $\chi$
with the lattice spacing, and a joint chiral/continuum extrapolation
was carried out for $0.08 \lesssim a \lesssim 0.12~\mathrm{fm}$ (over
a factor of 2 in $a^2$). The behaviour was as expected: the continuum
limit of $f_\pi = 91.6~(8.8)~\mathrm{MeV}$ was remarkably consistent
with physical expectations, with discretisation effects tending to
increase its value (as seen in other determinations
\cite{AliKhan:2001tx}).

The discretisation effects tended to increase the topological
susceptibility on all the ensembles. This agrees with expectations
for the chiral limit and is in contrast to the quenched theory, where
they suppress $\chi$. This suggests that, relative to the quenched
theory, there are fundamental differences in the vacuum due to sea
quarks, even for the largest $m_q$. The topological susceptibility is
more sensitive to these differences than, for example, most quantities
in the light hadron spectrum.

In summary, the topological susceptibility is a sensitive probe of the
vacuum and chiral properties of the lattice action. Although the
${\cal O}(a)$--improved SW action formally breaks the chiral symmetry
at finite lattice spacing, the behaviour of $\chi$ suggests that the
effects are not significant at the lattice spacings simulated. The
chiral behaviour of this action is comparably good to that of other
actions currently used in large--scale simulations, and makes it a
suitable laboratory for studying topologically sensitive states,
including the $\eta^\prime$ meson.

\begin{acknowledgments}
  
  The author is grateful to G.~Schierholz and M.~Teper for useful
  discussions; to D.~Hepburn, A.C.~Irving and D.~Pleiter for providing
  estimates of quantities for ensembles $u_5$ and $q_{15}$ in
  Table~\ref{tab_ensembles}; and to the Royal Society for financial
  support.

\end{acknowledgments}


\end{document}